\documentclass[reprint,
amsmath,amssymb,
aps,
prstab,
]{revtex4-1}

\usepackage{graphicx}
\usepackage{dcolumn}
\usepackage{bm}

\begin{document}


\title{Characterization of superconducting RF breakdown by two-mode excitation}

\author{G. Eremeev}
 \email{grigory@jlab.org}
\author{A.D. Palczewski}
 \affiliation{Thomas Jefferson National Accelerator Facility, Newport News, VA 23606, U.S.A.}

\date{\today}

\begin{abstract}
We show that thermal and magnetic contributions to the breakdown of superconductivity in radiofrequency (RF) fields can be separated by applying two RF modes simultaneously to a superconducting surface. We develop a simple model that illustrates how mode-mixing RF data can be related to properties of the superconductor. Within our model the data can be described by a single parameter, which can be derived either from RF or thermometry data. Our RF and thermometry data are in good agreement with the model. We propose to use mode-mixing technique to decouple thermal and magnetic effects on RF breakdown of superconductors.

\end{abstract}

\maketitle


\section{INTRODUCTION}
With the advances in superconducting radiofrequency (SRF) cavity fabrication and preparation over the last four decades, the peak surface magnetic field in superconducting niobium cavities approaches the critical field of niobium \cite{4441242},\cite{Furuta51}. The best SRF cavities lose their superconductivity at about $B\mathrm{_{peak}}$ = 180--200 mT, which is close to the thermodynamic critical field of niobium\cite{French1968}. Theoretical calculations suggest the scaling of the superheating field, which is believed to be the limitation of superconductivity in RF fields, to the thermodynamic critical field\cite{PhysRevB.85.054513},\cite{PhysRevB.83.094505},\cite{PhysRevB.78.224509}. Experimental data near the transition temperature supports the superheating critical field as the limitation in RF fields\cite{PhysRevLett.39.826}, but in experiments at low temperatures SRF cavities are often limited to fields lower than the expected superheating field, possibly due to local imperfections. How the cavity quench field relates to the material critical field is not evident, because RF dissipation increases the temperature of RF surface and magnetic and thermal effects on the superconducting RF breakdown become entangled.
\newline\indent We show that thermal and magnetic effect can be decoupled by exciting two RF modes. The idea was exploited in 1980 by D. Proch et al., guided by the intuitive understanding that magnetic breakdown and thermal breakdown will behave differently in time-varying fields\cite{1061065}: in a multicell SRF cavity the surface field distribution in each cell is the same among all resonances of a fundamental mode passband, but relative field strengths between different cells vary from mode to mode. If two pass-band modes in a cell are excited at the same time, their field vectors are collinear, but they do not add up coherently because of different resonant frequencies. The incoherent summation leads to beating in the amplitude of the resulting wave, where the maximum field is equal to the sum of field amplitudes, and the time averaged dot product of two fields is zero. The thermal breakdown will be determined by the sum of field amplitudes squared, while magnetic breakdown is the consequence of maximum field amplitude. D. Proch et al. studied $\pi$ and $\pi$/2 excitation in several 2-cell cavities. By bringing cavities into repetitive quench with varying field strengths ratio of two modes, quench dependence on field amplitude ratio was measured. The conclusion was that ``the data unambiguously supports the thermal model." 
\newline\indent Our data deviates from the quadratic law that we expected for thermal instability. We explain the data by magnetic breakdown on surface features with thermally suppressed critical field, and model how thermal and magnetic properties of the material result in the observed behavior. The experimental technique can be applied to characterize SRF niobium cavities as well as to characterize the RF breakdown of superconductivity in other superconducting devices.
\section{RF breakdown by two-mode excitation}
Superconducting properties of an SRF cavity are typically described by the curve showing the quality factor as a function of the RF field. The best superconducting cavities reach fields up to 200 mT with a decrease in the quality factor, so-called medium field Q-slope. Temperature mapping shows that temperature increase of the sensors' temperature over the superfluid helium bath temperature is roughly quadratic with the field in this regime\cite{EremeevDissertation}. The temperature increase of the sensor's temperature corresponds to the temperature increase on the RF surface of the cavity, so we will use quadratic approximation of the RF surface temperature:
\begin{equation}
T_\mathrm{RF}\cong T_{0}+A\cdot H_\mathrm{RF}^{2},
\end{equation}
where $T_{0}$ is the liquid helium bath temperature, and $A$ is the coefficient of proportionality. To understand how coefficient A depends on niobium wall properties, let us consider a simple model.
\newline\indent Following ~\cite{doi:10.1142/S1793626812300058}, let us consider a one-dimensional geometry, where a slab of niobium with thickness $d$ = 3 mm is cooled by a liquid helium bath at 2 K on one side, and on the other side a time-varying RF field $\vec{H}\mathrm{_{RF}}$ is applied in vacuum. The characteristic propagation time for a thermal wave in this geometry is $\tau_{T}~\approx~d^{2}C_{v}/\kappa~$\cite{Uroev}. Niobium volumetric heat capacity $C_{v}$ is $~\approx~10^{3}$ J/(K$\cdot~m^{3}$)\cite{PhysRev.109.788} and the thermal conductivity $\kappa$ is $~\approx~$10 W/(K$\cdot~m$) at 2 K, so $\tau_{T}~\approx~$1 msec. This time is much  smaller than RF field amplitude modulation times $Q_{L}$/$\omega~\cong~Q_{0}/\omega~\approx~$1 sec for a critically coupled cavity with a typical intrinsic quality factor $Q_{0}~\cong~10^{10}$, but it is much larger than RF period 1/f$~\cong~$ 1 nsec, so the heat flow equation can be reduced to steady state simultaneous equations:
\begin{equation}
\begin{split}
R_{s}(T_\mathrm{RF})\overline{\vec{H}_\mathrm{RF}^{2}}d&=\int^{T_\mathrm{RF}}_{T_{s}}\kappa(T)dT\\
R_{s}(T_\mathrm{RF})\overline{\vec{H}_\mathrm{RF}^{2}}&=H_\mathrm{Kapitza}(T_{s},T_{0})\cdot(T_{s}-T_{0}),
\end{split}
\end{equation}
where $R_{s}(T_\mathrm{RF})$ is the surface resistance, $\kappa(T)$ is the thermal conductivity of niobium, $H_\mathrm{Kapitza}(T_{s},T_{0})$ is the Kapitza conductance, $\overline{\vec{H}_\mathrm{RF}^{2}}$ is the time-averaged RF field squared, $T_\mathrm{0}$ is the helium bath temperature, $T_\mathrm{s}$ is the temperature of the niobium surface on the helium side, and $T_\mathrm{RF}$ is the temperature of RF side. The pre-heating temperature rise before a quench measured on the surface of a superconducting RF cavity is usually on the order of 0.1 K\cite{Gurevich200638,doi:10.1142/S1793626812300058}, so assuming $\kappa(T)\cong \kappa(T_\mathrm{0}) = \kappa$ and $H_\mathrm{Kapitza}(T_\mathrm{s}, T_\mathrm{0})\cong H_\mathrm{Kapitza}(T_\mathrm{0},T_\mathrm{0}) = H_\mathrm{Kapitza}$ the temperature on the RF surface can be expressed:
\begin{equation}
T_\mathrm{RF}\cong T_{0}+C\cdot R_\mathrm{s}\overline{\vec{H}_\mathrm{RF}^{2}},
\label{RFsurfaceT}
\end{equation}
where $C$ is $\left(\frac{d}{\kappa}+\frac{1}{H_\mathrm{Kapitza}}\right)$. 
\newline\indent When two modes are excited, the resulting field is the vector sum of two fields, $\vec{H}_\mathrm{RF}$ = $\vec{H}_{1}e^{i\omega_{1}t}$ + $\vec{H}_{2}e^{i\omega_{2}t}$, where $\vec{H}_{1}$, $\vec{H}_{2}$ field amplitudes are parallel for TM$_\mathrm{010}$ passband modes. Since the thermal diffusion time is much larger than $2\pi$/$\omega_{1}$, $2\pi$/$\omega_{2}$, and $2\pi$/$(|\omega_{1}-\omega_{2}|~\approx$ 1 $\mu$sec, averaging the RF field over the thermal diffusion time results in $\overline{\vec{H}_\mathrm{RF}^{2}}$ = $\frac{1}{2}(H_{1}^2+H_{2}^2)$. Hence, when two fields are applied:
\begin{equation}
T_\mathrm{RF}\cong T_{0}+C\frac{1}{2}R_{s}(H_{1}^2+H_{2}^2)
\label{TwofieldT}
\end{equation}
\newline\indent The result illustrates that the characteristic times for a typical SRF cavity are such that the total dissipation on the RF surface can be expressed as the sum of losses. The result supports the intuitive speculation that the power losses of two passband modes in an SRF cavity add up independently.
\newline\indent For magnetic breakdown we expect niobium to go from superconducting- to normal-conducting state, when the applied RF magnetic field exceeds the RF critical field on the surface. \textit{We will assume that the magnetic breakdown occurs when the amplitude of magnetic field exceeds the RF critical field, that is in the case of mixed modes we assume that the breakdown occurs when the sum of RF field amplitudes exceeds the RF critical field}, i.e., $H_{1}+H_{2}=H_\mathrm{crit}^\mathrm{RF}(T)$. The temperature dependence of $H_\mathrm{crit}^\mathrm{RF}(T)$ is still under discussion. The measurements of T. Hays et al.\cite{HaysHsh} as well as more recent ones by N. Valles et al.\cite{VallesHsh} are fitted with the accepted approximation for the thermodynamic critical field, while earlier measurements of T. Yogi\cite{PhysRevLett.39.826} show a divergence of H$^\mathrm{RF}_{c}$/H$_{c}$ near T$_{c}$. Recent theoretical calculations\cite{PhysRevB.85.054513},\cite{PhysRevB.83.094505},\cite{PhysRevB.78.224509} show non-trivial dependence of H$^\mathrm{RF}_{c}$/H$_{c}$ within different approximations. Since there is no universally accepted RF critical field approximation for niobium at 2 K with the Ginsburg-Landau parameter $\cong$ 1, \textit{we will use the textbook formula for the temperature dependence of the thermodynamic critical field~\cite{ashcroft} to model the temperature dependence of RF critical field}:
\begin{equation}
H_{c}(T_{RF})=H_{c}\left(1-\left(\frac{T_{RF}}{T_{C}}\right)^2\right),
\label{Hrffielddep}
\end{equation}
where $H_{c}$ is $H_{c}(0)$, the critical field at zero temperature, $T_{c}$ is the critical temperature, and $T_{RF}$ is the temperature of the RF surface. When two RF fields are applied, the sum of the field amplitudes of two modes is equal to the critical magnetic field:
\begin{equation}
H_{1}+H_{2}=H_{c}\left(1-\left(\frac{T_{0}+C\frac{1}{2}R_{s}(H_{1}^2+H_{2}^2)}{T_{C}}\right)^2\right),
\label{twofields}
\end{equation}
where $T_{c}$ and $H_{c}$ are the critical temperature and the magnetic field, and we used ($\ref{RFsurfaceT}$) for RF surface temperature. When only one field is applied:
\begin{equation}
H_\mathrm{max}=H_{c}\left(1-\left(\frac{T_{0}+C\frac{1}{2}R_{s}H_\mathrm{max}^2}{T_{C}}\right)^2\right),
\label{onefield}
\end{equation}
where $H_\mathrm{max}$ is the breakdown field amplitude. Equations ($\ref{twofields}$) and ($\ref{onefield}$) can be reduced to:
\begin{equation}
\begin{split}
&\left(\frac{H_{1}}{H_\mathrm{max}}+\frac{H_{2}}{H_\mathrm{max}}\right) + 
\\& + \frac{R_{s}H_{c}T_{0}H_\mathrm{max}C}{T_{c}^2}\times
~\left(\left(\frac{H_{1}}{H_\mathrm{max}}\right)^2+ \left(\frac{H_{2}}{H_\mathrm{max}}\right)^2\right)+
\\&+\frac{R_{s}^2H_{c}H_\mathrm{max}^3C^2}{4T_{c}^2}\times
~\left(\left(\frac{H_{1}}{H_\mathrm{max}}\right)^2+ \left(\frac{H_{2}}{H_\mathrm{max}}\right)^2\right)^2 =
\\&= 1+\frac{R_{s}H_{c}T_{0}H_\mathrm{max}C}{T_{c}^2}+\frac{R_{s}^2H_{c}H_\mathrm{max}^3C^2}{4T_{c}^2}
\end{split}
\label{modemixingwithoutalpha}
\end{equation}
If we define 1/$\alpha$=$1+\frac{R_{s}H_{c}T_{0}H_\mathrm{max}C}{T_{c}^2}+\frac{R_{s}^2H_{c}H_\mathrm{max}^3C^2}{4T_{c}^2}$ and normalize the fields $H_{1}$ and $H_{2}$ to H$_{max}$, the resulting equation can be written in terms of $\alpha$, $T_{0}$, $T_{c}$ and normalized fields $\tilde{H}_{1}$ and $\tilde{H}_{2}$ as:
\begin{equation}
\begin{split}
&\alpha(\tilde{H}_{1}+\tilde{H}_{2})+
\\& + 2\frac{T_{0}^2}{T_{c}^2-T_{0}^2}\left(\sqrt{\frac{T_{c}^2}{T_{0}^2}(1-\alpha)+\alpha} -1\right)(\tilde{H}_{1}^{2}+\tilde{H}_{2}^{2})+
\\&+\frac{T_{0}^2}{T_{c}^2-T_{0}^2}\left(\sqrt{\frac{T_{c}^2}{T_{0}^2}(1-\alpha)+\alpha} -1\right)^2(\tilde{H}_{1}^{2}+\tilde{H}_{2}^{2})^2=1
\end{split}
\label{modemixingbestfit}
\end{equation}
\newline\indent To understand intuitively the physical meaning of $\alpha$, we will assume that $\alpha$ is close to 1, that is $\alpha=1-\delta$, where $\delta<<1$. Expanding equation ($\ref{modemixingbestfit}$) and dropping quadratic and higher order terms in $\delta$, we obtain the following expression for mode-mixing:
\begin{equation}
\alpha(\tilde{H}_{1}+\tilde{H}_{2})+(1-\alpha)(\tilde{H}_{1}^{2}+\tilde{H}_{2}^{2}) = 1
\label{modemixingbestfitsimple}
\end{equation}
From this equation we can see that $\alpha=1$ means a purely magnetic quench. As $\alpha$ departs from one, the thermal effects become more pronounced.
\newline\indent To understand the physical meaning of $\alpha$ quantitatively, we expand the equation ($\ref{twofields}$) and, after combining with equation ($\ref{modemixingwithoutalpha}$) and the definition of $\alpha$, the fitting parameter $\alpha$ can be re-written as:
\begin{equation}
\alpha=\frac{H_\mathrm{max}}{H_{c}\left(1-\left(\frac{T_{0}}{T_{C}}\right)^2\right)}
\label{alphaphyssense}
\end{equation}
This expression illustrates the physical meaning of the fitting parameter $\alpha$. The fitting parameter quantifies how much the material critical field is suppressed due to RF surface temperature increase over the liquid helium bath temperature. In analogy with geometrical field enhancement factor, $\alpha$ can be called \textit{the thermal suppression factor}.
\newline\indent Equation ($\ref{modemixingbestfitsimple}$) clearly shows that pre-quench temperature rise varies with $\alpha$. In the case of $\alpha$ = 0, the pre-quench temperature rise is the same for all mixed field amplitudes; in the case of $\alpha$ = 1, the pre-quench temperature rise dependence on one of the fields is parabolic with $\tilde{H}_\mathrm{min}$ = 0.5 and $\Delta T_\mathrm{min}/\Delta T_\mathrm{max}$ = 0.5. Here $\Delta T_\mathrm{max}$ is the maximum pre-heating temperature rise during mode mixing, $\Delta T_\mathrm{min}$ is the minimum pre-heating temperature rise, and $\tilde{H}_\mathrm{min}$ is the normalized field level at which the minimum temperature rise occurs. 
\section{EXPERIMENTAL SETUP}
All studied cavities followed the current standard the International Linear Collider (ILC) cavity preparation procedure\cite{GengCrawfordILCstandard}. A general overview of modern preparation techniques can be found in [\onlinecite{SurfPrepTTC2008}]. Quality factor vs. field for $\pi$ and other TM$_\mathrm{010}$ pass-band cavity modes was measured using conventional phase-lock loop techniques; see e.g.,[\onlinecite{PadamseeHaysKnobloch}]. All RF tests were done in the liquid helium bath at 2 K. Helium temperature was kept constant by cryosystem feedback and was monitored by a calibrated temperature sensor in the helium bath. For the pass-band measurements the transmitted power from the pick-up probe was used to calculate the RF field amplitude in the end cells; the RF fields for the other cells were calculated using the end cell amplitude\cite{WangPassbands}.
\newline\indent During measurements, second sound propagation in the liquid helium bath\cite{ConwayOST} and temperature mapping techniques\cite{ILC2CellJlabtmap} were used to identify quench locations, and to confirm that breakdown happens in the same location during the mode-mixing measurement. All the pass-band modes of a cavity are measured first and the highest field in each cell for each mode is calculated. Quench locations are determined with oscillating superleak transducers (OSTs) and the limiting cell for each pass-band mode is identified. After OSTs and thermometry data have been analyzed, two modes that are limited by the same defect are chosen for mode-mixing measurements.
\newline\indent Once the pass-band modes are chosen, we use two independent voltage-controlled oscillators (VCOs) with independent phase-lock loops (PLL) to drive modes independent of each other. Each VCO is tuned to one of the modes. The drive signals from VCOs are combined with a power combiner and fed into the high power RF amplifier. From the power amplifier through RF cables and the directional coupler, RF power from two sources is fed into the cavity via the input coupler of the cavity. Phase locking is accomplished by splitting transmitted power from the pick-up probe with power dividers and feeding the transmitted signal into the PLL of both VCOs. Part of the transmitted power is coupled into the input port of the spectrum analyzer. The spectrum analyzer is then used to measure field level of each mode during mode-mixing experiments. More details about setup, measurement technique, and calibration can be found elsewhere\cite{DualmodeSRF11}.
\newline\indent Measured mode-mixing data is normalized to the quench field in each mode and presented on an x-y plot. Each data point on the x-y plot corresponds to a single RF breakdown event during mode-mixing. The abscissa of the point is equal to one of the mixed fields normalized amplitudes; the ordinate of the point is equal to the other field normalized amplitude. As the result an x-y dependence connecting points (0,1) and (1,0) is plotted. To fit the experimental data we used: 
\begin{equation}
\begin{split}
&\alpha(\tilde{H}_{1}+\tilde{H}_{2})+
\\& + 2\frac{T_{0}^2}{T_{c}^2-T_{0}^2}\left(\sqrt{\frac{T_{c}^2}{T_{0}^2}(1-\alpha)+\alpha} -1\right)(\tilde{H}_{1}^{2}+\tilde{H}_{2}^{2})+
\\&+\frac{T_{0}^2}{T_{c}^2-T_{0}^2}\left(\sqrt{\frac{T_{c}^2}{T_{0}^2}(1-\alpha)+\alpha} -1\right)^2(\tilde{H}_{1}^{2}+\tilde{H}_{2}^{2})^2=1,
\end{split}
\end{equation}
where $\alpha$ is a fitting parameter and $\tilde{H}_{1}$ and $\tilde{H}_{2}$ are the field amplitudes of each mode normalized to the maximum field amplitude measured with single-mode excitation. The critical temperature $T_{c}$ is set to 9.25 K, and the bath temperature $T_{0}$ is set to 2 K.
\newline\indent Thermometry is commonly used in an SRF field to measure dissipation distribution on the cavity surface. The temperature measured by a temperature sensor on the outside surface of a cavity is proportional to the temperature on the inside surface. It is commonly accepted that for Allen-Bradley thermometry\cite{KnoblochPhD}, which is widely used in SRF fields, the measured temperature is proportional to the power flux flowing though the sensor from the RF surface into the liquid helium bath:
\begin{equation}
\Delta T \propto P^\mathrm{RF} = \overline{R_{s} (\vec{H_{1}}+\vec{H_{2}})^{2}} = \frac{1}{2} R_{s} (H_{1}^{2} + H_{2}^{2})
\label{TempRise}
\end{equation}
Our thermometry system was not designed for the calibrated temperature measurements of the RF surface temperature; it measures the maximum temperature rise before quench with the thermometry sensor located on the outside surface in the helium bath. We show the thermometry experimental results here, to illustrate that the temperature data qualitatively agrees with RF mode-mixing data.
\section{EXPERIMENTAL RESULTS AND DISCUSSION}
\begin{figure}[htb]
   \centering
   \includegraphics*[width=80mm]{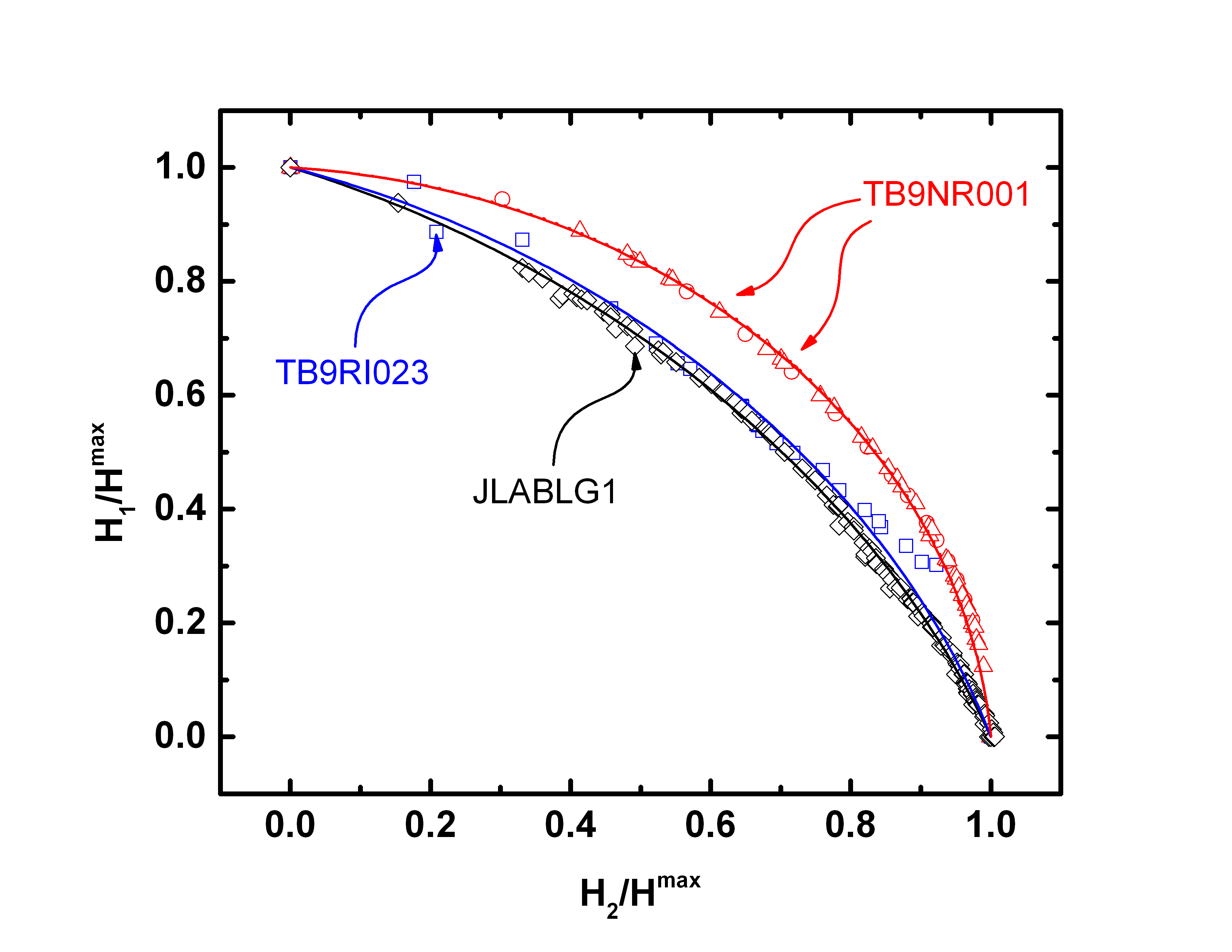}
   \vspace{-.1in}
   \caption{RF mode-mixing measurements of TB9NR001, TB9RI023, and JLab-LG-1. Open blue squares present the data for 6$\pi$/9 and 5$\pi$/9 mode mixing in TB9RI023; the blue solid line present the best fit using ($\ref{modemixingbestfit}$) with the best fit $\alpha$ of 0.58 $\pm$ 0.02. Open red circles and triangles present the data for $\pi$ $\&$ 7$\pi$/9 and 3$\pi$/9 $\&$ 7$\pi$/9 of TB9NR001 mode mixing; the best fits are presented with solid and dot red lines; the best fits for $\alpha$ are 0.20 $\pm$ 0.01 and 0.19 $\pm$ 0.01 respectively. JLab-LG-1 $\pi$ $\&$ 6$\pi$/9 mode-mixing results are shown with black rhombs. The best fit $\alpha$ = 0.63 $\pm$ 0.00 is shown with the solid black line.}
   \label{BothRFmeasurements}
\end{figure}

\begin{figure}[!htb]
   \centering
   \includegraphics*[width=80mm]{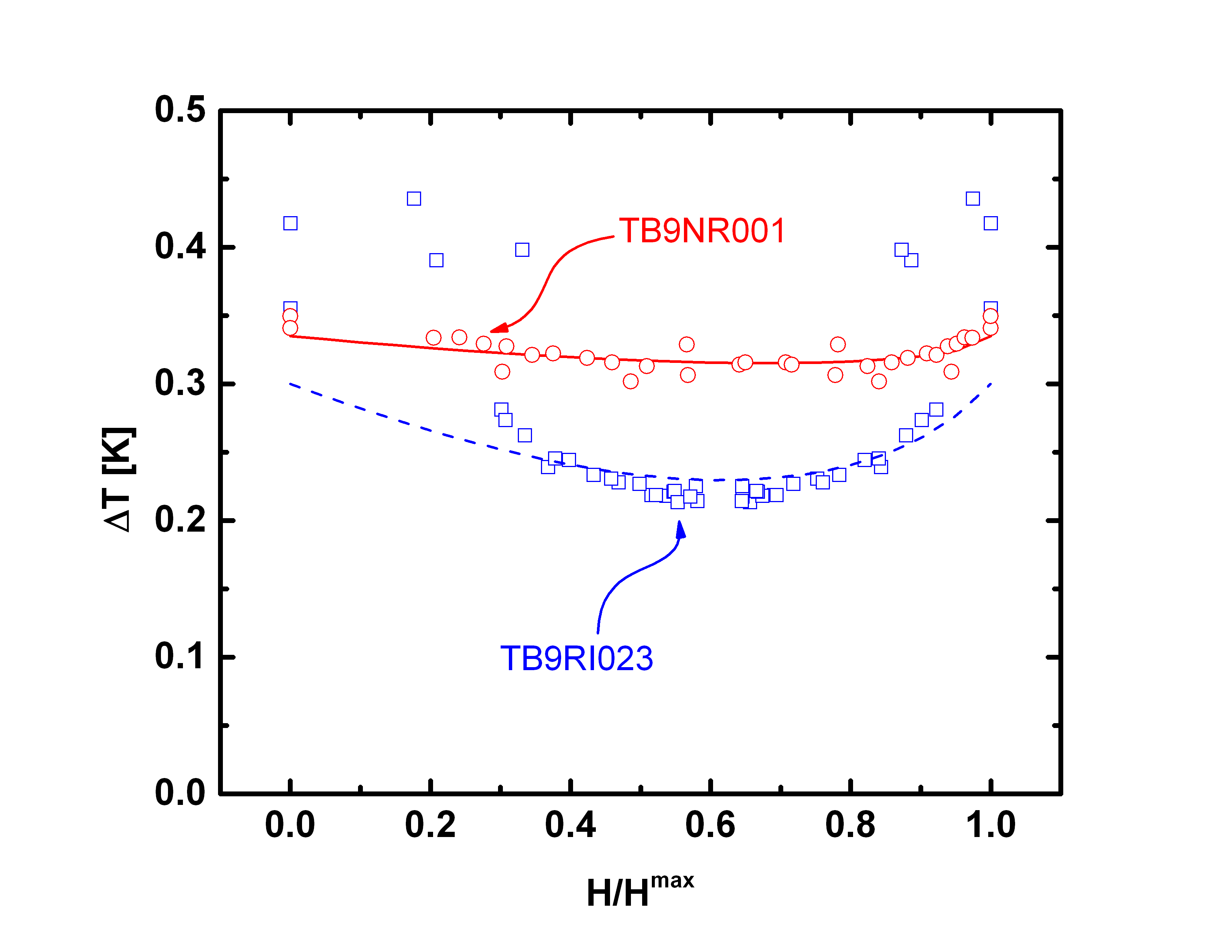}
   \vspace{-.1in}
   \caption{Temperature rise as a function of field during mode-mixing measurement. Blue open squares present the data for TB9RI023; the dashed blue line is the expected dependence with $\alpha$ = 0.58; $\Delta$T$_\mathrm{max}$ is set to 0.3 K. Red open circles are the experimental measurement for TB9NR001; the solid red line is the expected dependence with $\alpha$ = 0.2; $\Delta$T$_\mathrm{max}$ is set to 0.335 K.}
   \label{ThermometryModeMixing}
\end{figure}

The RF measurement results for cavities TB9NR001, TB9RI023, JLAB-LG-1 are shown in Fig.$\ref{BothRFmeasurements}$. TB9NR001 and TB9RI023 were produced by two industrial vendors for the ILC R$\&$D yield studies. Both cavities are fine grain ($\approx$ 50 $\mu$m) 9-cell ILC cavities made of high RRR ($\approx$ 300) bulk niobium. JLab-LG-1 is an R$\&$D 9-cell cavity manufactured at JLab from high RRR large grain ($\approx$ 5 cm) material\cite{KneiselLG1}. The standard RF qualification measurement consists of locking TM$_\mathrm{010}$ $\pi$-mode resonant frequency and measuring quality factor, while step-wise ramping up the field in approximately B$_{peak}~\cong$ 5 mT increments up to the quench field. TB9NR001 was limited in $\pi$ mode to B$_{peak}$ =  70$\pm$4 mT. Thermometry and OSTs measurements indicated that the cavity is limited at the equator in the fifth cell. Quenches in all odd pass-band modes were confirmed with OSTs to happen at the same location. For dual mode excitation measurements $\pi$($f_\mathrm{TM_{010}}^{\pi}~\cong$ 1.29985 GHz) and 7$\pi$/9($f_\mathrm{TM_{010}}^{7\pi/9}~\cong$ 1.29664 GHz) modes were mixed first, later during the same cool down 3$\pi$/9($f_\mathrm{TM_{010}}^{3\pi/9}~\cong$ 1.27981 GHz) and 7$\pi$/9 modes were mixed. TB9RI023 in 6$\pi$/9, 5$\pi$/9, and 2$\pi$/9 modes was limited by a defect in the third cell. 6$\pi$/9($f_\mathrm{TM_{010}}^{6\pi/9}~\cong$ 1.29313 GHz) and 5$\pi$/9($f_\mathrm{TM_{010}}^{5\pi/9}~\cong$ 1.28908 GHz) were used for mode-mixing measurements, because these modes are better coupled to the cavity by the fixed coupler that we use. Following the same experimental procedure as with TB9NR001, the RF and thermometry data was recorded.
\begin{table}
\caption{In this table we summarize cavity IDs, passbands, measured end cell peak fields, and limitations for the passbands used during mode mixing measurements. Note that the listed peak magnetic field are the peak magnetic field measured in the end cell and calculated\cite{WangPassbands} peak magnetic field in the limiting cell.}
\begin{tabular}{| c | c | c | c | c |}
  \hline                       
 	 Cavity ID & mode & B$_{peak}^{max}$ [mT] & B$_{peak}^{max}$ [mT] & limit\\
 	 & & (end cell) & (limiting cell) & \\
  \hline                       
 	 TB9NR001 & $\pi$ & 70 & 70 & cell $\#$5\\
 	 & & & & quench\\
  \hline                       
 	 TB9NR001 & 7$\pi$/9 & 72 & 74& cell $\#$5\\
   & & & & quench\\
  \hline                       
 	 TB9NR001 & 3$\pi$/9 & 35 & 69 & cell $\#$5\\
   & & & & quench\\
  \hline                       
 	 TB9RI023 & $\pi$ & 72 &  72 & Field\\
 	 & & & & emission\\
  \hline                       
 	 TB9RI023 & 6$\pi$/9 & 149 & 146 & cell $\#$3\\
   & & & & quench\\
  \hline                       
 	 TB9RI023 & 5$\pi$/9 & 115 & 139 & cell $\#$3\\
   & & & & quench\\
  \hline                       
 	 JLab-LG-1 & $\pi$ & 86 & 86 & cell $\#$6\\
   & & & & quench\\
  \hline                       
 	 JLab-LG-1 & 6$\pi$/9 & 103 & 101 & cell $\#$6\\
   & & & & quench\\
  \hline                       
\end{tabular}
\label{SummaryTable}
\end{table}
In Fig.$~\ref{BothRFmeasurements}$ the results from dual mode excitation measurements of $\pi$ $\&$ 7$\pi$/9 and 3$\pi$/9 $\&$ 7$\pi$/9 modes along with the best fits are presented. The TB9NR001 measurements best fit for $\pi$ $\&$ 7$\pi$/9 is 0.20 $\pm$ 0.01, the best fit for 3$\pi$/9 $\&$ 7$\pi$/9 is 0.19 $\pm$ 0.01. The TB9RI023 measurements best fit for 5$\pi$/9 $\&$ 6$\pi$/9 is 0.58 $\pm$ 0.02. JLab-LG-1 was limited in $\pi$ mode by a defect in the sixth cell at B$_\mathrm{peak}$ = 86 $\pm$ 5 mT. The same limitation was encountered in 6$\pi$/9, 4$\pi$/9, and $\pi$/9 modes. The results for $\pi$($f_\mathrm{TM_{010}}^{\pi}~\cong$ 1.30054 GHz) and 6$\pi$/9($f_\mathrm{TM_{010}}^{\pi}~\cong$ 1.29365 GHz) mode-mixing are shown in Fig.$\ref{BothRFmeasurements}$. The best fit yields 0.63 $\pm$ 0.01.
\newline\indent One of the Allen-Bradley thermometers that was located at the identified quench location sites and showed the highest temperature rise before the quench was used for dual-mode thermometry measurements. The thermometer was disconnected from 2-cell thermometry DAQ multiplexer and connected to a separate 8-channel DAQ. The voltage across the thermometer along with the transmitted power was then recorded as a function of time for different mode-mixing amplitudes. The highest voltage before the quench for different mode-mixing amplitudes was identified from the transmitted power and the thermometer voltage trace. The voltage was then converted to the temperature rise with the calibration curve. The temperature rise data points as a function of one of the mixed fields is presented in Fig.$~\ref{ThermometryModeMixing}$ for TB9NR001 and TB9RI023. During TB9RI023 thermometry measurements we observed sharp rise in temperature after the quench for $\tilde{H}$ close to 0 or 1, but did not see the sharp temperature rise for $\tilde{H}$ between 0.3 and 0.9.  At this time it is unclear why this changed occurred, but it was reproducible and could be a future research topic outside the scope of this paper. 
\newline\indent As we pointed earlier, the thermometry is not calibrated against the RF surface temperature, so we did not try to fit the data with theoretical expression. Instead, we calculated the expected temperature rise for the best fit $\alpha$ from RF measurements. In Fig.$~\ref{ThermometryModeMixing}$ the experimental thermometry data points with the scaled overlay of fitted curves from RF data shows strong agreement between two techniques. As expected for TB9NR001 from RF measurement with an $\alpha~\approx$ 0.2  the temperature shows little variation, whereas TB9RI023 data with RF $\alpha~\cong$ 0.58 shows the expected dip at about half-field value.
\newline\indent Using the equation ($\ref{alphaphyssense}$), the quench fields from the table $\ref{SummaryTable}$, and the fitting parameter $\alpha$ we can infer the material critical field at the quench site for the three quenches. From the quench field in TB9NR001 of $\approx$ 70 mT and $\alpha$ = 0.2, we infer the material critical field in TB9NR001 is $\approx$ 350 mT. This value is about 50 percent higher than 240 mT at 2 K typically assumed to be the limiting superheating field of niobium\cite{French1968},\cite{Matricon1967241}. We speculate that the discrepancy stems from the assumption that the breakdown is magnetic, whereas in TB9NR001 the quench is likely to be a thermal quench from a normal conducting inclusion\cite{Highrestwindefect}. If the quench in TB9NR001 is in fact thermal then equation ($\ref{twofields}$) does not describe the breakdown condition and equation ($\ref{alphaphyssense}$) is not applicable. From the quench field in TB9RI023 of $\approx$ 140 mT and $\alpha$ = 0.58, we infer the material critical field in TB9RI023 is $\approx$ 241 mT. This value compares well with 240 mT for the generally accepted value of the superheating field. We note that in the case of TB9RI023 the quench site was off the equator, and therefore the derived quench field, which is the calculated peak magnetic field in the cell, is overestimated. From the quench field in JLAB LG-1 of $\approx$ 86 mT and $\alpha$ = 0.63, we infer the material critical field in JLAB LG-1 is $\approx$ 137 mT. This value is about 40 percent lower than 240 mT, which can be understood, if we take into account a geometrical field enhancement factor of the large grain cavity. We note that in JLAB LG-1 a sharp feature was found at the quench location\cite{WatanabeDefectSRF2011}.
\section{CONCLUSION}We have studied breakdown of superconductivity in several multicell superconducting radio frequency cavities by simultaneous excitation of two TM$_\mathrm{010}$ pass-band modes. The experimental data is characterized by one parameter, which is related to geometry and material properties of the superconductor. Unlike measurements done in the past, which indicated a clear thermal nature of the breakdown, our measurements at higher fields present a more complex picture with interplay of both thermal and magnetic effects. The limiting defects were characterized with RF measurement and thermometry, which we have shown are independent of each other. Mode-mixing parameters calculated from RF and thermometry data show a fair agreement for both measured defects. Either of these measurement techniques, RF or thermometry, can be used to quantify the superconducting RF breakdown.
\begin{acknowledgments}
We would like to thank for suggestions and useful discussions Gigi Ciovati, Rongli Geng, Charlie Reece, Bob Rimmer, and Hui Tian. We are grateful to Jefferson Lab SRF staff for helping with cavity preparation and assistance with RF tests.
\newline\indent This manuscript has been authored by Jefferson Science Associates, LLC under Contract No. DE-AC05-06OR23177 with the U.S. Department of Energy. The United States Government retains and the publisher, by accepting the article for publication, acknowledges that the United States Government retains a non-exclusive, paid-up, irrevocable, world-wide license to publish or reproduce the published form of this manuscript, or allow others to do so, for United States Government purposes.
\end{acknowledgments}

\bibliographystyle{unsrt}

\bibliography{myreferences} 

\end{document}